\documentclass[letterpaper,11pt]{article}

{\mathsc}{OT1}{cmr}{m}{sc}

\newcommand{\SM}       {\mathsc{sm}}
\newcommand{\Planck}   {\mathsc{pl}}

\newcommand{\GUT}      {\mathsc{gut}}
\newcommand{\TeV}      {~\mathrm{TeV}}
\newcommand{\GeV}      {~\mathrm{GeV}}
\newcommand{\hc}       {\mathrm{\; h.c. \;}}
\newcommand{\order}{\mathcal{O}}
\newcommand{\gappeq}{\mathrel{\rlap {\raise.5ex\hbox{$>$}}
{\lower.5ex\hbox{$\sim$}}}}

\begin{document}

\begin{center}
            \hfill    UPR-1133-T\\
\end{center}

\begin{center}
{\large \bf Twenty-five Questions for String Theorists}
\end{center}

\begin{center}
P.~Bin\'etruy$^{1}$, G.~L.~Kane$^{2}$, J.~Lykken$^{3}$ and
B.~D.~Nelson$^{4}$
\end{center}

\begin{center}
$^{1}${\it LPT, Universit\'e Paris-Sud, Bat. 210,\\

91405 Orsay CEDEX, France}
\end{center}

\begin{center}
$^{2}${\it Michigan Center for Theoretical Physics, Randall
Laboratory,\\

University of Michigan, Ann Arbor, MI 48109}
\end{center}

\begin{center}
$^{3}${\it Theoretical Physics Department,\\

Fermi National Accelerator Laboratory, Batavia, IL 60510}
\end{center}

\begin{center}
$^{4}${\it Department of Physics \& Astronomy,\\

University of Pennsylvania, Philadelphia, PA 19103}
\end{center}

\begin{quotation}
In an effort to promote communication between the formal and
phenomenological branches of the high-energy theory community, we
provide a description of some important issues in supersymmetric
and string phenomenology. We describe each within the context of
string constructions, illustrating them with specific examples
where applicable. Each topic culminates in a set of questions that
we believe are amenable to direct consideration by string
theorists, and whose answers we think could help connect string
theory and phenomenology.
\end{quotation}

\section*{Introduction}
Today there is a renewed interest in phenomenology among some part
of the string theory community, this despite the fact that
sometimes string theorists state that string theory is too poorly
understood to try to connect it to the real world. While we would
not disagree with that, we think that the needed understanding is
more likely to emerge if string theorists take up the study of
certain phenomenologically relevant questions. We are hopeful that
proposing several questions can encourage string theorists to
solve them, or at least formulate them more precisely, and thereby
generate closer connections between phenomenological theory and
string theory.

The relevance of many of the phenomenological issues we choose to
consider will be obvious to even the most formal of theorists,
though the relevance of some others may not be as clear. We have
chosen to introduce each question (or set of questions) with a
discussion of the low-energy physics issues that motivate it. We
then attempt to embed this discussion in string theory, often by
means of examples from both older and more recent constructions.
As we are often reminded by string theorists, the exact nature of
string theory is not yet fully known. Thus a series of examples
may not indicate a theorem about all of M-theory -- or even about
a particular class of constructions. But we hope they will serve
to illustrate the reason for our interest in these topics and
stimulate thinking in a phenomenological direction more generally.

The questions fall naturally into several groups. We have chosen
to list them in order of increasing apparent difficulty. By this
we mean that for some questions the avenue of approach from string
theory is somewhat clear, while for others it is more obscure. The
most difficult questions will be those that pertain most directly
to the low and intermediate-energy world, in particular those that
involve some level of dynamical symmetry breaking. It is often
said that such field-theory dynamics is by definition not the
subject of study for string theorists. But such thinking is
counter-productive: the dynamics of a field theory are often {\em
determined} by the boundary conditions we apply at the string
energy scale, and these boundary conditions are most certainly in
the domain of string theory.

There are of course a number of additional questions that can be
posed beyond the ones contained here, and we encourage others to
suggest them. When questions overlap, but are logically distinct,
we have not tried to consolidate them; redundancy might be
helpful. Furthermore, there are areas that one would call
``phenomenological'' that formal theorists are actively pursuing:
obtaining $N=1$ supersymmetry or three generations in the
low-energy theory, attacking the cosmological constant problem,
issues of early universe physics, moduli stabilization, and so on.
These are crucial topics, but as our goal is to stimulate thinking
in directions that have been somewhat overlooked, we will not
consider them here.

Partial answers to some questions are already known to some
people. But these answers are generally not known in forms useful
to promote phenomenological progress. We urge experts to publicize
results that broaden understanding along the lines we discuss, and
in additional areas not covered here as well.

Finally, we do not intend to write a review of open issues in
string phenomenology, nor of supersymmetric phenomenology more
generally.\footnote{Indeed, an excellent one already exists, in a
format similar to this one~\cite{Dienes:1997nq}.} Thus, after some
thought we have tried to provide only a brief discussion, with
limited references to work that can be considered relevant to the
questions, so as to not dilute the impact of a short, precise
paper. Those who answer the questions will provide the references.
Much of what follows will be familiar to many, but many reviews of
supersymmetric and string phenomenology are available for further
elucidation~\cite{Ellis:1998eh,Dudas:2000bn,Munoz:2003au,Chung:2003fi,Abel:2004rp}.

\section{Gauge symmetries}
\label{sec:gauge}
%

\subsection{Large groups}
\label{sec:rank}

A basic question of phenomenological importance is the rank of the
gauge group that results in four dimensions from compactification
of the string theory. Symmetries have historically been both our
primary means of modeling known phenomena as well as our primary
tool for building theories of the unknown. From the string
perspective we expect many such symmetries that are operative in
the low-energy world to have an origin as a gauge symmetry. The
set of models and mechanisms which rely on some new extension of
the Standard Model gauge group is large, and utilizing several in
a complete model can quickly put strong demands on the size of the
gauge group descending from string theory.

As an example, we might start with the Standard Model and its
rank-four gauge group $G_{\SM} = SU(3) \times SU(2) \times U(1)$.
Achieving a group of this size is quite elementary for most string
constructions. But phenomenologists often add much more to this
basic structure. For example, a gauged $U(1)_{\rm B-L}$ can be
extremely useful in insuring an intact R-parity, and hence the
necessary stability of the proton and/or a cold dark matter
candidate. Similarly, if we wish to generate an effective
$\mu$-term from the vev of some field we may need yet another
$U(1)$ factor. And if we wish to generate Majorana masses for
neutrinos we may need yet another. We might also wish to introduce
so-called ``horizontal'' or ``family'' symmetries that operate on
the generation indices of the Standard Model fields in order to
generate certain Yukawa coupling patterns. While the simplest of
these could be just a single $U(1)$ factor, more realistic models
often call for higher rank flavor symmetries, either as products
of $U(1)$ factors or as non-Abelian groups such as $SU(2)$ or
$SU(3)$~\cite{Dine:1993su,Nir:1996am,Babu:1999js,Ross:2002fb}.
Thus it is not unimaginable that a gauge group with rank seven,
eight or even larger might be the best candidate for describing
the Standard Model degrees of freedom.

On top of this we must consider the issue of supersymmetry
breakdown. It is natural to consider gaugino condensation as a
candidate mechanism in the string theory context. While in
principle any asymptotically free gauge group will suffice
(notably {\em not~} one of the groups we mentioned in the previous
paragraph), in practice achieving the right scale of SUSY breaking
in the observable sector often requires a group of fairly
substantial size, such as $SU(5)$ or
higher~\cite{Gaillard:1999et}. If we wish to use only the
tree-level K\"ahler potential for the dilaton we must employ
multiple condensates in the hidden sector with beta-functions that
are tuned against one another. Examples of such configurations
with combined rank less than eight are rare. Indeed, the typical
rank of the combined condensing group often needs to be
$\mathcal{O}(10)$ -- even larger than sixteen in many
cases~\cite{deCarlos:1992da}.

One might argue that accommodating all of this rank is not a
problem in string models. But specific string constructions often
involve a limit on the allowed rank of the low-energy gauge group.
For example, standard constructions of the weakly coupled
heterotic string (WCHS) give rise to gauge groups at the (4D)
string energy whose rank can be no larger than 16. If one were to
allow some or all of the compact space-time coordinates to be
associated with the gauge group, as in the free-fermionic or
asymmetric orbifold constructions, this rank could (in principle)
be as high as 22 at the string scale. However, it could also be
much lower: reduction in rank between the higher-dimensional
theory and the 4D theory can occur in compactification through the
vacuum values of continuous Wilson lines which act as adjoint
Higgs representations. And realizing the gauge group at higher
affine level in the underlying conformal theory results in a
reduction in the rank of the gauge group as seen from the 4D
theory.

This tension between phenomenological demands and the mathematical
consistency of the string theory is not confined to the heterotic
region of the M-theory space. Semi-realistic (and supersymmetric)
models of particle physics involving Type~I and Type~II string
theory have thus far been constructed mainly on toroidal orbifolds
or orientifolds, typically involving intersecting D-branes. In
these cases mathematical consistency of the theory restricts the
number of such branes -- and hence the rank of the gauge group(s)
that can appear on them -- to 16 or less. It is certainly possible
to modify these conditions so as to achieve larger rank, perhaps
arbitrarily large. But to date no such construction with a
semi-realistic gauge group and spectrum exists without abandoning
supersymmetry in the low-energy theory. While the demands of
phenomenologists on such models continue to push for larger
rank,\footnote{For example, the ``racetrack inflation'' model
based on Type~IIB string theory of~\cite{Blanco-Pillado:2004ns}
wanted two asymptotically free gauge groups with ranks of
$\order(10)$, and possibly as large as $\order(100)$.} these
models may (in fact) be even more constraining than the WCHS.
Achieving sufficient intersection numbers to obtain three families
of chiral matter often restricts the allowed rank of the resulting
gauge group. Including non-trivial magnetic flux generally
restricts the rank even more -- perhaps fatally by eliminating
candidate hidden sector gauge groups for gaugino
condensation~\cite{Marchesano:2004yq}. Recent surveys of the
string theory landscape for flux compactifications of Type~IIB
string theory suggest that small rank gauge groups are, indeed,
generally preferred~\cite{Kumar:2004pv,Conlon:2004ds}. Faced with
this sort of evidence we pose the question:

\vspace{0.2in}

\noindent \textbf{Question 1:} What are the properties of string
constructions that can provide realistic observable sectors while
simultaneously providing large-rank gauge groups in a hidden
sector?\\

Where specific mechanisms for alleviating this tension between
mathematical consistency and phenomenological necessity have
already been suggested (for example, the inclusion of
anti-D-branes, as in~\cite{Cascales:2003zp,Marchesano:2004xz}) we
would further ask: \\

\noindent \textbf{Question 2:} Can general relations between the
mechanism(s) that relax restrictions on the rank of the low-energy
gauge group, and the overall low-energy phenomenology of the
construction, be identified?

\subsection{Gauge coupling unification}
\label{sec:unification}

It has now been more than twenty years since the remarkable
observation was
made~\cite{Dimopoulos:1981yj,Dimopoulos:1981zb,Ibanez:1981yh} that
if (a) the hypercharge generator is given a normalization
consistent with its inclusion in an $SU(5)$ group, (b) the
Standard Model field content is extended to a supersymmetric model
in the minimal manner, (c) these superpartners have masses near
the electroweak scale, and (d) there are no additional fields in
the theory charged under the Standard Model gauge group to some
high-enough energy scale, then the renormalization group (RG)
evolution of the measured gauge couplings of the Standard Model
within this paradigm leads to a unification of the three couplings
at a very high energy scale to a very good degree of accuracy. The
intervening twenty years of measurements of Standard Model
phenomena have done nothing to weaken this stunning fact -- in
fact, more precise data have only made the conclusion
stronger~\cite{deBoer:2003xm}. Hence the heavy focus of the
theoretical community on low-energy supersymmetry.

While it has become somewhat fashionable of late to take the
implicit (if not explicit) view that this apparent unification is
merely a coincidence, a more careful study shows that it would
have to be a very great coincidence indeed. Even including the
uncertainty in superpartner masses (but keeping them within a
window near $1\TeV$), the locus of points in the
$\alpha_3(M_Z)\;-\;\sin^{2}\theta_w(M_Z)$ plane consistent with
unification at {\em any} scale is very small.
The fact that the measured values of these two parameters fall
within this locus of points was estimated by the authors
of~\cite{Ghilencea:2001qq} to have a chance of occurring in the
(0.2-2)\% range if it were merely a coincidence.

It has often been said that gauge coupling unification is a
``prediction'' of string theory, but this is somewhat misleading.
While it is true that eventually all coupling constants can be
related to the string coupling $g_{s}$ and string tension
$\alpha'$, that does not imply that the 4D theory just below the
string scale exhibits a unification of 4D gauge couplings. The
relation between the (unified) string coupling and the various
low-energy couplings appearing in the effective field theory in
four dimensions involves the vacuum expectation value of various
moduli associated with the string compactification. Even in the
WCHS, where this relation is universal for all gauge groups at the
leading order, threshold corrections can affect this leading-order
unification and possibly spoil the ``prediction'' of gauge
coupling unification.

Despite this caveat the compatability of the WCHS with the {\em
idea~} of high-scale gauge coupling unification was held as a
major phenomenological success of this theory. Yet in recent
approaches where the Standard Model is generated by overlapping
and/or intersecting D-brane configurations gauge coupling
unification is not at all automatic unless the gauge groups of the
Standard Model are all embedded on one stack of D-branes. And just
as in the WCHS case, threshold corrections from heavy string modes
will affect the various couplings of the four-dimensional gauge
groups differently, unless a grand unified theory with a simple
GUT group exists below the string scale. These threshold
corrections involve a host of moduli associated with the
compactification itself and will take different forms depending on
the type of theory.

Given the apparent challenges for even the WCHS in explaining the
apparent unification of gauge couplings at a scale $\Lambda_{\GUT}
\sim 10^{16} \GeV$ -- let alone achieving the correct
normalization of hypercharge and the absence of exotic states that
ruin the RG evolution~\cite{Dienes:1996du} -- we might first well
ask the question:\\

\noindent \textbf{Question 3:} If the apparent gauge coupling
unification is not a coincidence, is this alone evidence for the
existence of a unification-scale GUT? \\

One would immediately think the answer is negative, given the fact
that plenty of examples can be constructed that do not involve
GUTs. For example, it was shown in~\cite{Nilles:1997vk} that
heterotic string theory allows for unification at the string scale
provided certain geometrical moduli and Wilson line moduli acquire
particular vevs, that need not be large. But Question~3 is asking
whether gauge coupling unification, at the scale and to the
accuracy inferred from low-energy measurements, could ever be
considered a {\em natural} or {\em generic} outcome in a string
model without an operative GUT theory below the string scale and
apart from any special configuration of moduli vevs. That is:\\

\noindent \textbf{Question 4:} What are the stringy conditions
that would guarantee gauge coupling unification occurs rather than
just imposing it? \\

Thus even if there is no grand unification into a simple group the
gauge coupling unification could still be natural if the sizes of
the moduli expectation values are about the same or related by
some symmetry. The latter may be the only hope that brane
constructions can be made compatible with gauge coupling
unification. A particularly valiant effort in this direction was
that of~\cite{Blumenhagen:2003jy} in which the geometrical
properties of a particular Type~IIA orientifold construction with
$D_6$-branes were exploited, along with an assumption about
certain moduli expectation values (namely, that the volumes of
certain 3-cycles that these branes wrap are equal) to argue that
gauge couplings should unify. Even here only one of the two
conditions needed for unification becomes satisfied. It is not
clear what could imply the second condition. Certainly the
challenge of explaining gauge coupling unification in open string
models will require such an interplay of compactification
geometry, brane configuration and symmetries among moduli. What we
seek is an explanation that can span several classes of brane
constructions, so we ask:\\

\noindent \textbf{Question 5:} What are the necessary or
sufficient conditions that brane constructions must have for
automatic gauge coupling unification to be the result? \noindent

\subsection{Hypercharge normalization}

A related, but logically distinct, question that impinges on gauge
coupling unification is the issue of hypercharge normalization.
Let us take a well-known example to illustrate the problem. In the
weakly coupled heterotic string with gauge group $E_8 \bigotimes
E_8$, compactified on an orbifold, we expect nonzero Wilson lines
to break the gauge group of the underlying string theory to a
product of gauge groups, including a number of $U(1)$ factors. It
is a highly nontrivial task to decide which linear combination of
these $U(1)$ factors should be identified as the hypercharge of
the Standard Model (see Section~\ref{sec:meaning} below). In fact,
the process of isolating the trace anomaly over $U(1)$ factors to
one linear combination generally implies mixing among Abelian
factors from the observable $E_8$ and the hidden $E_8$ group,
obscuring any residual knowledge of the underlying $E_6$ or SO(10)
structure. Finally, cancellation of the Fayet-Iliopoulos D-term
associated with the anomalous $U(1)$ generally results in the
breaking of one or more additional $U(1)$ factors, suggesting that
identifying the correct hypercharge combination can only be done
after model-dependent field theory dynamics occurs. Once this
process is carried out, it is typically the case that the
hypercharge normalization relative to the other generators of the
Standard Model gauge group is not that which would result from a
grand unified group, i.e. $k_Y \neq 5/3$.

We have illustrated the issue of hypercharge normalization with an
example of the heterotic string on a toroidal orbifold, but
intersecting brane configurations in open string theories are no
less problematic. One may hope to understand a normalization
consistent with $SU(5)$ unification in the context of the WCHS
since the standard embedding results in an $E_6$ gauge factor
prior to the introduction of Wilson lines. But in $D$-brane
constructions the starting gauge group is generally a product of
$U(N)$ factors, without any point in the construction at which a
unified group could be said to exist. While the normalization of
the Abelian factor in the decomposition $U(N) \to SU(N) \times
U(1)$ may be clear, the one factor of such $U(1)$'s identified as
hypercharge emerges as an arbitrary linear combination -- the GUT
normalization is then understandable only as a coincidence. While
the situation improves somewhat in cases where an underlying
Pati-Salam $U(4)\times U(2)_L \times U(2)_R$ structure is
employed, the fact remains that due to the lack of universal gauge
coupling determination (see Section~\ref{sec:unification} above),
hypercharge normalization must still be seen as a fortuitous
accident of moduli dynamics. An easier solution is to seek out
models that guarantee $k_Y = 5/3$.\\

\noindent \textbf{Question 6:} What are the necessary or
sufficient conditions that would guarantee the normalization $k_Y
= 5/3$ for the $U(1)$ factor associated with hypercharge in each
class of string constructions?\\

Even in the absence of such a mechanism the observed high-energy
unification of coupling constants could be the result of
additional matter with the appropriate charges introduced at
particular intermediate scales.~\footnote{For a discussion of
this, and other issue of gauge unification in string models, we
refer the reader to~\cite{Dienes:1996du} and references therein.}
In this sense it is possible that the so-called ``exotics,'' which
are usually considered a nuisance could turn out to be a virtue.
Thus we might seek a second-best solution:\\

\noindent \textbf{Question 7:} Are there any reasons to believe
that particular compactifications might automatically produce
extra states with the right properties to maintain gauge coupling
unification even with $k_Y \neq 5/3$?

\subsection{Gauge singlets as matter fields}

One of the most often-employed tools of the model-builder is the
gauge singlet: by eliminating all possible gauge interactions it
becomes easier to solve certain problems in isolation. Examples
include the inflaton and curvatons of early universe cosmology,
fields to generate neutrino Majorana masses (indeed often the
neutrinos themselves -- see below), fields to generate
$\mu$-terms, fields to drive phase transitions in the early
universe, etc. Some of these solutions work less efficiently if
these fields carry quantum numbers for some other gauge groups
beyond those of the Standard Model.

In string theory, if we classify all 4D fields as either moduli or
matter fields, and then agree to define a field as ``matter'' if
it carries quantum numbers under some gauge interactions, then by
implication all such fields described above would have to be
moduli. This common rule of thumb about matter and gauge
interactions is supported by the dimensional reduction of gauge
interactions from ten dimensional supergravity to four dimensions.
Matter states and their superpotential couplings in 4D are then
seen as remnants of the 10D super-Yang-Mills interaction, and
hence always involve gauge-charged matter.

But the term ``modulus'' has a more precise technical meaning as a
field with no classical potential -- a field which parameterizes
the degeneracy of the vacuum of the string theory. Many examples
of moduli that are familiar to string model builders fill both the
technical definition (no classical superpotential couplings) and
the common definition (no gauge interactions). These include the
moduli that arise from dimensional reduction of the tensors
$g_{MN}$ and $b_{MN}$ of ten-dimensional supergravity, and the
dilatonic scalar. Yet many other fields which also parameterize
the degeneracy of vacua in a particular string construction do not
obey one or the other of the two definitions given above. For
example the degrees of freedom that parameterize continuous Wilson
lines can be represented as gauge-charged
matter~\cite{Ibanez:1987pj,Mohaupt:1993fb,Cvetic:2000aq}. In
toroidal orbifolds states with a non-zero oscillator number and
gauge interactions can be thought of as would-be blowing-up modes
that parameterize the transition from a singular (orbifold)
manifold to a smooth (Calabi-Yau)
one~\cite{Font:1988tp,Cvetic:1987bi,Cvetic:1999qx}.

In both of these cases the above-mentioned gauge-charged fields
have a second role that is moduli-like in nature (i.e. they
parameterize a property of the compactification geometry) and
typically have superpotential couplings. But fields without gauge
charges that might otherwise be unambiguously called moduli can
behave as matter as well. Consider, for example, the case of
vector-bundle moduli in heterotic string theory compactified on a
Calabi-Yau threefold. The elements of $H_1({\rm End} V)$ for the
bundle $V$ with structure group $G$ will be singlets of the gauge
group that is the complement to $G$ in $E_8$. For example, if $V$
is the $SU(3)$ tangent bundle in the standard embedding, the
elements of $H_1({\rm End} T)$ will be singlets of $E_6$. The
speculation that these fields might couple linearly to
gauge-charged $\mathbf{27}$ and $\overline{\mathbf{27}}$'s in the
superpotential was put forward some time ago~\cite{Witten:1985bz}.
However, in the standard embedding the vector-bundle moduli and
geometrical moduli are identified, and any possible superpotential
couplings would be severely constrained. More recent work that
goes beyond the standard embedding in constructions with $G =
SU(4)$ has shown that the vector-bundle moduli, now liberated from
their geometric brethren, do indeed couple linearly in the
superpotential as has been demonstrated by explicit
construction~\cite{Braun:2005bw}. Yet even the relatively mundane
geometrical moduli that characterize the gross shape of the
compact space can be thought of as having polynomial
superpotential couplings to gauge-charged matter if one is in a
regime of moduli space where a suitable expansion of the
moduli-dependent Yukawa couplings is available.

Thus it seems that the distinction between ``matter'' and
``moduli'' is somewhat blurred. Perhaps an inappropriate use of
language is to blame, but the issue of true gauge singlets having
polynomial superpotential couplings to gauge-charged matter is
more serious than mere semantics. If singlet fields couple to
charged matter in such a way as to be relevant to low-energy
phenomenology then they may fill the roles mentioned at the
beginning of this section. But employing true singlets for such
tasks as generating Majorana masses for neutrinos or generating a
low-energy Higgs bilinear often results in the breaking of
unwanted global symmetries with accompanying cosmological
difficulties. And singlets with superpotential couplings to
Standard Model fields can destabilize the hierarchy between the
electroweak and string scales when supergravity loop effects are
considered~\cite{Bagger:1993ji}. Thus it is of great utility to
have an unambiguous definition of matter fields vis-a-vis moduli
fields in string terms, or put differently:\\

\noindent \textbf{Question 8:} Under what conditions can true
singlets of all gauge symmetries exist and in what sense can they
be truly called ``matter'' (i.e. have Yukawa interactions with SM
fields)?\\

As an interesting corollary, let us note that model builders often
assume that right-handed neutrinos, introduced to generate
neutrino masses, are true singlets themselves. This is made
plausible by the fact that they are singlets under the Standard
Model gauge group. But as mentioned above, since we wish these
fields to have Yukawa couplings to left-handed doublets it seems
unlikely that they can be overall gauge singlets. Nevertheless, if
they were we might expect mixing between neutrinos and various
string moduli, and such ideas have been suggested in the
past~\cite{Benakli:1997iu}. In theories with open-string states a
true singlet neutrino might prompt the question of whether
neutrinos were in fact open string states or closed string states.
Given the difficulty in finding natural manifestations of Majorana
mass terms for neutrinos in string models (see
Section~\ref{sec:neutrino} below) we might wonder if the unique
nature of the neutrino in the Standard Model particle content is
somehow a reflection of something radical:\\

\noindent \textbf{Question 9:} If right-handed neutrinos are true
singlets, do they mix with string moduli?\\

\section{Discrete symmetries}
\label{sec:discrete}
%

\subsection{Classifying electroweak doublets}

In the Standard Model the combination $B-L$, where $B$ is the
baryon charge of a field and $L$ is its lepton charge, is a
perturbatively conserved global quantum number. When the Standard
Model gauge group arises from the decomposition of $SO(10)$ or
$E_6$ then $B-L$ can be promoted to a local gauge symmetry and is
generated by a linear combination of the various Abelian
generators of the decomposition. Interestingly, $B-L$ is one of
only two linearly independent additional $U(1)$ factors that can
be added to the Standard Model and be anomaly-free with just the
Standard Model field content. Most importantly, an exactly
conserved $B-L$ implies a perfectly stable proton. Thus this
particular $U(1)$ factor -- or a discrete subgroup of it -- holds
an important place in low energy phenomenology.

In this section we wish to focus on lepton number, returning to
baryon number below. Most semi-realistic string constructions with
$N=1$ supersymmetry, whether of the heterotic or open-string
variety, have a spectrum with more fields than those of the MSSM.
These ``exotics'' often contain $SU(2)$ doublets that are $SU(3)$
singlets. Even when an underlying $SO(10)$ structure was once
present, the need for twisted sectors in the superstring spectrum
often implies that the states of the low-energy spectrum need not
fill out complete $\mathbf{16}$ and $\mathbf{10}$ representations.
Thus an immediate ambiguity ensues: should these exotic fields be
classified as Higgs doublets or lepton doublets?

Without $SO(10)$ to guide us, or a gauged $U(1)_{\rm B-L}$ or
discrete R-parity which can be embedded in an $SO(10)$ structure,
the only way to distinguish between lepton and Higgs doublets is
through their superpotential couplings to other quarks and
leptons. Yet the notion of lepton doublet versus Higgs doublet may
be meaningless from the point of view of string selection rules.
While some degree of mixing between the two species of doublet may
be tenable for low-energy phenomenology, arbitrary mixing
generally is not~\cite{Banks:1995by}. Most importantly, proper
electroweak symmetry breaking and the absence of massless
Pecci-Quinn axions requires a TeV-scale bilinear for just one type
of doublets -- those that we identify as Higgs states.

If we must appeal solely to string selection rules and the string
physics which determines Yukawa interactions in order to
distinguish leptons from Higgs fields, then the structure of the
MSSM Lagrangian can only be a fortuitous accident in the sense
that not all possible operators are to be allowed. In theories of
intersecting $D$-branes the analog to sectors at various fixed
points is the identifying of MSSM states with strings localized at
the intersection of various branes containing Standard Model gauge
groups. Then $U(1)_B$ and $U(1)_L$ might {\em individually~} be
understood as arising from the decomposition $U(3) \to SU(3)
\times U(1)_B$ and $U(2) \to SU(2) \times U(1)_L$. That is,
different sectors of the Hilbert space, determined by geography in
the compact space, are correlated with representations under the
SM gauge groups, with particle number symmetries understood
topologically. This is an appealing way to understand these
symmetries in the absence of $SO(10)$ gauge symmetry, but does not
fully distinguish leptons from Higgs states -- not to mention the
problem of generating a Higgs bilinear (see
Section~\ref{sec:muterm} below). Thus a fundamental concern of
low-energy phenomenology is the following:\\

\noindent \textbf{Question 10:} Can a definition of lepton number
that distinguishes lepton doublets from Higgs doublets be
unambiguously defined for string theory in the absence of an
underlying SO(10) gauge structure?

\subsection{Proton decay and baryon number violation}

Experimental searches for proton decay will eventually probe
lifetimes of about $10^{35}$ years. The longevity of the proton
suggests that higher-dimensional operators formed solely from
Standard Model fields must be suppressed by a large mass scale
($\gappeq 10^{16} \GeV$). For supersymmetric theories with a high
string scale ({\em i.e.} near the Planck scale) such as the ones
we imagine here, this requires that all operators of mass
dimension four and five which can mediate proton decay must be
eliminated from the theory, or at least adequately suppressed.
Some set of discrete symmetries is generally assumed to accomplish
this task. Of course, if baryon number is related to some gauged
symmetry, such as $U(1)_{\rm B-L}$ then such operators can be
forbidden if this symmetry is broken only spontaneously by the
appropriate vev of some field.

As mentioned above, $D$-brane models provide a framework for
understanding the gauging of baryon number via the emergence of
the Standard Model $SU(3)$ factor from a fundamental $U(3)$
factor. But $D$-brane constructions also tend to involve
fundamental string scales that are somewhat or significantly below
the scale $M_{\Planck} = 2 \times 10^{18} \GeV$. In many cases a
spontaneously broken gauged $U(1)_B$ is not enough to protect the
proton -- some additional discrete symmetry is again needed. It is
almost always the case in heterotic models (even those such as
free-fermionic constructions based on the NAHE basis
set~\cite{Faraggi:1996yu,Ellis:1997ec} where a ``frustrated''
$SO(10)$ symmetry can help forbid some operators in the
superpotential) that forbidding dimension four and five operators
require particular discrete symmetries.

We will have more to say about these symmetries in the following
question, but here our interest is in the dimension six operators,
assuming that lower dimensional ones are somehow forbidden to
all-orders. Generally speaking, operators that mediate proton
decay at dimension six arise from the exchange of new gauge bosons
that connect quarks and leptons. In many string constructions,
particularly heterotic ones, a unified group containing such gauge
fields exists {\em prior~} to the imposition of Wilson line
breaking upon compactification. As such, these states are
projected out of the light spectrum and (for large string scales)
are incapable of inducing proton decay observable in any
forseeable experiment. This conclusion, though drawn after
consideration of heterotic models, continues to hold in open
string constructions such as Type~IIA theory with
$D_6$-branes~\cite{Klebanov:2003my}. What is most interesting,
from a phenomenological point of view, is whether the converse
holds, namely:\\

\noindent \textbf{Question 11:} In a theory where dimension four
and five proton decay operators are forbidden, would the
observation of proton decay imply the unification (in four
dimensions) of quarks and leptons in a simple gauge group, or
could observable proton decay arise in such a string theory
without grand unification?

\subsection{R-parity}

A related but not identical question is whether there are
mechanisms that can be discerned from the string theory level to
guarantee the presence of a matter parity, or R-parity, in the
low-energy superpotential. Such an R-parity would do much (but not
all) to resolve the issues in the previous question. Furthermore,
in the presence of such a parity the lightest supersymmetric
particle (LSP) is stable. The stability of the LSP is important
enough to warrant its own question (see below), so here we focus
specifically on the issue of an intact matter parity such as
R-parity.

An intact R-parity without a gauged $U(1)_{\rm B-L}$ generally
arises as an accident in string constructions, if at all. To
determine if an accidental R-parity exists, a laborious procedure
is required. After calculating the spectrum of light states one
must compute the allowed Yukawa interactions and then identify the
states of the Standard Model via their leading Yukawa interactions
and allowed hypercharge assignments. After this is done, and
possible F- and D-flat directions are identified, the effective
superpotential terms that survive along each flat direction can be
studied to determine the existence of discrete symmetries such as
R-parity. Since the Standard Model gauge group does not forbid
R-parity violating terms, constructions that lead directly to the
Standard Model may have such terms. Some may arise at the leading
(trilinear) order, and others may arise at higher order in an
expansion in $M_{\Planck}^{-1}$, presumably reflecting
higher-genus terms from the underlying string theory.

These higher-order terms cannot be neglected, for if R-parity is
violated, the magnitude of violation must be very small. While
certain R-parity violating operators can exist in isolation with
$\order(1)$ coefficients, as argued above string models that do
not have an intact R-parity tend to allow all such operators at
various mass-dimensions in the superpotential. A more satisfying
solution is to identify R-parity as a symmetry of the underlying
compactification geometry -- such as a remnant of a modular
symmetry~\cite{Gaillard:2004aa}. As the geometry of the
compactified space determines the superpotential (both the
tree-level and higher-genus terms) it is reasonable to ask the
following question:\\

\noindent \textbf{Question 12:} Would any such R-parity be an
exact symmetry of the string theory or could it be an approximate
parity? If the latter, how large might the violations be?

\subsection{A stable new state}

The longevity of the LSP has profound implications, both for
collider phenomenology and for cosmology. In collider events a
long-lived LSP, which is presumed to be electrically neutral,
exits the detector and the missing transverse energy can be used
as a trigger. Many of the most familiar supersymmetry search
strategies rely on this trigger for the analysis. Here it is the
relevance of a stable LSP for cosmology that we wish to consider.

In cosmology, a stable LSP provides a possible weakly-interacting
dark matter candidate -- though the details of whether such a
particle can indeed be the missing non-baryonic dark matter is a
model-dependent issue. In contrast, if the LSP has a finite
lifetime then it is unlikely to provide the required dark matter.
Here we are explicitly talking of the lightest state in the
superpartner spectrum and the discrete symmetry protecting its
lifetime is presumably the R-parity of the previous question.
Other stable or quasi-stable particles could emerge in the string
theory spectrum and have been suggested in this context:
modulinos~\cite{deCarlos:1993jw}, exotic gauge-charged
matter~\cite{Chang:1996vw}, hidden sector matter
composites~\cite{Benakli:1998ut}, hidden sector gauge
composites~\cite{Faraggi:2000pv} and wrapped
D-branes~\cite{Shiu:2003ta}. It could be that one or more of these
states (or others not yet imagined) contribute to the dark matter
of the cosmos and thus it is appropriate to ask about the
stability of new states in a general context:\\

\noindent \textbf{Question 13:} What string theory conditions are
sufficient to guarantee stable states beyond the Standard Model
particles?

\section{Flavor and CP}
\label{sec:flavor}
%

\subsection{The meaning of flavor}
\label{sec:meaning}

The problem of understanding the physics behind the masses and
mixings of the Standard Model fermions is a venerable one which
has yet to be satisfactorily solved. Yet many approaches to a
solution exist, some of which have already been alluded to (see
Section~\ref{sec:rank} above). Phenomenologists try to understand
the hierarchies in fermion masses through symmetries, either
discrete or continuous, that act on some flavor index that
represents the three states of each species of field in the
Standard Model~\cite{Kane:2005va}. Thus it might at first seem
that translating these symmetries to the string level is a matter
of simply understanding how the triple replication of families
arises in string theory. We will have more to say about this
translation process in the next question, but there is something
more fundamental to be considered first: just what does ``flavor''
mean from the string point of view? Let us illustrate the thought
process behind this question with an example from heterotic string
theory.

When working out the massless spectrum of the bosonic heterotic
string on an orbifold we ``tag'' different fields by a whole host
of properties that are hidden from view in the four-dimensional
effective field theory. Here are some examples: untwisted states
will carry some amount of H-momentum. Twisted states will be
identified by fixed point locations -- on the $Z_3$ orbifold this
means identifying a fixed point location in each of three possible
complex planes. Furthermore, there can be twisted fields with
non-zero (left-moving or right-moving) oscillator number of some
amount, and the direction of this oscillator excitation in the
compact space must also be specified. These ``internal quantum
numbers,'' if you will, determine the Yukawa couplings through the
string selection rules, so they are ``proto-flavor'' in nature.

Thus there is presumably, for every string construction that gives
rise to three generations, a natural basis in which to embed the
concept of proto-flavor. In some constructions this natural basis
is easy to identify.  For example, in the $Z_3$ orbifold of the
bosonic heterotic string it is natural to obtain three generations
in the following manner. One postulates the existence of a
non-trivial Wilson line in the first two of the three complex
planes of the factorizable orbifold. In each of the three complex
planes the $Z_3$ action leaves three points invariant. A twisted
sector state is then labeled in part by its geography: the fixed
point location in each of the three planes. Since there are three
possibilities in each plane, one naively expects a 27-fold
redundancy in the spectrum, or 27 generations of matter. But the
presence of Wilson lines in two of the three planes brings this
number to three. Then we might say that (modulo issues of
oscillator number) ``proto-flavor'' -- that is, generation number
labels -- are naturally defined by fixed point location in the
third complex plane.

This may not be completely satisfactory to explain the actual
flavor structure of the Standard Model (see the next question) but
it is a natural starting point. In other constructions the path to
three-fold replication is less transparent, so the embedding of
generation indices in the string quantum numbers is more
complicated~\cite{Chaudhuri:1995ve}. Nevertheless a natural
embedding should be identifiable. For example, in the
free-fermionic models based on the NAHE basis set there are
nominally 48~generations, but additional projections reduce this
to one per basis vector
$b_i$~\cite{Antoniadis:1989zy,Faraggi:1992fa,Faraggi:1993pr}.
These $b_i$ are boundary conditions for the free fermionic fields,
specifying the phase they acquire after parallel transport around
a noncontractable loop. In the analogous $Z_2 \times Z_2$
asymmetric orbifold construction these three generations are
identified with the three possible twist assignments under the two
$Z_2$ factors.

In open string theories involving intersecting $D$-branes the
multiplicity of states is given by the intersection numbers of the
cycles wrapped by branes on which these open strings end. Here it
is not necessary that each state in the low-energy spectrum
receive the same multiplicity. Furthermore, explicit examples may
have two ``generations'' arising from different intersections from
the third -- see, for example, the construction
in~\cite{Cvetic:2002qa}. A rough analogy can be made, however,
between states at the intersection of branes at angles and twisted
states of heterotic orbifolds, and indeed just such a $2+1$
splitting was engineered in a recent $Z_6$ symmetric heterotic
orbifold~\cite{Kobayashi:2004ya}. So once again, the
classification of compactifications by their geometrical
properties should also give rise to a classification of
``proto-flavor'' symmetries, perhaps associated with
representations under various twist symmetries, prompting us to
ask:\\

\noindent \textbf{Question 14:} Can constructions be classified by
the manner in which generation number is embedded in the
string-theoretic properties of the light spectra?

\subsection{Selection rules as flavor symmetries}

Let us assume the answer to Question~14 is affirmative and a
classification of some set of string models is carried out. What
we would have for the models in this set is a set of symmetry
relations that act on some properly defined generation index at an
energy scale just below the compactification scale. This is not
yet a true theory of flavor, but it is a suggestive starting
point. But unless each generation of matter fields in the MSSM,
which fill out complete multiplets of $SU(5)$ (or $SO(10)$ if we
include the right-handed neutrino), come from the {\em same~}
sector of the string Hilbert space there is no reason to believe
that the embedding described above represents what we mean by
flavor in the low-energy sense. That is because at low-energies we
define flavor by the Yukawa couplings of the light fields. Thus,
at some level, the proto-flavor of the string theory must be
translated into a theory of flavor as reflected in the labels
``up, down, strange,...'' of the Standard Model.

What is more, the spectrum that survives to the electroweak scale
(and which presumably includes the MSSM particle content) may be
quite different from that which exists just below the string
scale. In general we expect vacuum expectation values to arise for
many fields along various flat directions. Typically this is
considered a welcome feature: it may allow for dynamically
generated mass terms that project out unwanted ``exotics'' from
the spectrum while Higgsing unwanted gauge symmetries. But typical
string models have many more states with Standard Model quantum
numbers than those of the MSSM. In the $Z_3$ example mentioned
above typical models may have $\order(10)$ anti-triplets of SU(3)
and $\order(20)$ SU(2) doublets. The ones that survive to be
labeled $d_i^c$, $u_i^c$, $L_i$, $H_u$ and $H_d$ are likely to be
linear combinations of states which remain massless along a given
flat direction and those that get large masses along the same
direction. In such a world the translation of ``proto-flavor''
symmetries into flavor symmetries involves a non-linear mapping.

Should we conclude from the above discussion that any flavor
symmetries which might descend from the underlying string theory
are likely to be discrete symmetries, or at best Abelian
continuous ones? To date these are the only such symmetries that
have been identified in explicit string constructions. Such a view
point might be supported by considering string selection rules
which serve to determine the allowed Yukawa interactions of light
fields. Such rules are relations that reflect the proto-flavor and
ultimately have a geometrical meaning, thus it should be possible
to understand them topologically. Their manifestation in the
effective field theory just below the string scale is typically in
the form of discrete symmetries. So we might first wish to know
the answer to the following question:\\

\noindent \textbf{Question 15:} How can string selection rules
which determine the superpotential be interpreted as low-energy
flavor symmetries?\\

And then, given that the answer to the above is likely to be
easiest for the case of discrete or continuous Abelian symmetries,
we ask the obvious follow-on question:\\

\noindent \textbf{Question 16:} Under what circumstances will
string-derived flavor symmetries take the form of continuous
non-Abelian horizontal symmetries when acting on the low-energy
degrees of freedom?

\subsection{Classification of K\"ahler manifolds}

The K\"ahler potential is every bit as important as the
superpotential in understanding the flavor structure of the
low-energy Lagrangian. Both at the tree level and the loop level
the elements of the curvature tensor formed by the K\"ahler
potential determine the flavor structure of soft terms and Yukawa
couplings, the latter through field-redefinitions necessary to
achieve a canonical kinetic energy term for the matter fields.
Isometries of the K\"ahler manifold for the non-linear sigma model
describing the light chiral superfields thus have some relevance
as flavor symmetries.

The supergravity models typically employed to study the
phenomenology of string theories start with a K\"ahler potential
for the matter fields that is diagonal in form
\begin{equation}
K = f(Z,\overline{Z})_{i} \left| \phi_{i} \right|^{2} + \dots,
\label{Kahler1} \end{equation}
where the fields $Z$ and $\overline{Z}$ represent some set of the
string moduli. This is true at least to the leading order
calculated from the string theory.\footnote{To fully understand
the flavor structure of the effective Lagrangian at string
energies it is necessary to know the complete curvature tensor.
This requires knowing at a minimum the higher-genus terms of the
K\"ahler potential trilinear and quartic in the chiral
superfields.} The ability to put the leading terms in the K\"ahler
potential into a diagonal form may result from the application of
certain isometries(in particular employing modular symmetries) and
presumably this diagonal form is in the basis defined by the
stringy quantum numbers we called ``proto-flavor'' above. Among
these string-defined properties are the representation of the
fields under the various gauge groups that result from the
compactification. For example, in the heterotic string theory on
an orbifold the untwisted sector of the theory exhibits such a
diagonal structure. Thus it may be natural to expect a form such
as
\begin{equation}
K = f(Z,\overline{Z})_{Q} \left| Q \right|^{2} +
f(Z,\overline{Z})_{U} \left| u^{c} \right|^{2} + \dots
\label{Kahler2} \end{equation}
to emerge.

Of course we are suppressing the generation, or flavor, indices on
these expressions. Even if we had a stringy definition of flavor
and it happened to be the case that each individual term in the
K\"ahler potential~(\ref{Kahler2}) was also diagonal in the
(low-energy) flavor basis we must still sum over many such terms
which appear at the same order, not to mention terms which appear
at higher order in the string theory. These terms, involving
fields with the same Standard Model quantum numbers, would tend to
involve a mixing of flavors that would be incompatible with the
diagonal assumption of the K\"ahler potential. Any such mismatch
between the stringy ``diagonal'' K\"ahler basis and the low-energy
flavor basis would suggest that flavor physics is at least in part
arising from the mismatch of fields in the superpotential and the
K\"ahler potential.
%
%
Again, a topological classification of K\"ahler manifolds plus a
stringy definition of flavor may shed light on this issue. It is
this rough idea that we might have in mind when we ask the
question: \\

\noindent \textbf{Question 17:} In what basis should we expect the
leading-order K\"ahler potential for massless gauge-charged fields
to be diagonal? Under what circumstances might this be a basis
that is also diagonal in the low-energy flavor basis?\\

But merely being diagonal is not completely sufficient to be safe
from all constraints from flavor-changing neutral current data,
though it is a major step in that direction. It is also necessary
for the {\em entries~} on the diagonal of the scalar mass
matrices, which depend on the K\"ahler metric, to be roughly equal
(in the basis in which fermion masses are diagonal). In other
words, if these entries are not equal then the super-CKM rotations
of the squarks to the quark basis will introduce flavor changing
effects proportional to these differences. Thus we need
universality in the functions $f(Z,\overline{Z})$
of~(\ref{Kahler1}) as well:\\

\noindent \textbf{Question 18:} If the K\"ahler metric is diagonal
in some field basis, under what circumstances should the values of
the diagonal entries be equal?

\subsection{Phases and CP}

When addressing the relation between string theory and complex
phases the most commonly taken strategy is to assume that CP is
violated spontaneously at the field theory level through the
(complex) vacuum expectation value of one or more string moduli.
Given a concrete moduli stabilization mechanism the values of
these complex vevs can presumably be deduced. As the couplings of
the various moduli to observable sector fields is known, this
should provide a way of understanding the nature of CP violation
in the CKM matrix and/or the soft supersymmetry-breaking
Lagrangian.

The ubiquity of this approach is often said to be motivated by the
assertion that CP is a gauge symmetry from the point of view of
the string theory and is preserved, perturbatively and even
nonperturbatively~\cite{Dine:1992ya}. Thus it is left to
spontaneous breaking of CP, capable of being described in an
effective field-theoretic manner, to account for the CP violation
of the observable world. But surely the string theory is not
completely mute on this point: we know that symmetry breaking that
may appear spontaneous in one context often appears geometric, or
explicit, in another context. For example, the study of orbifolds
with discrete torsion seems to indicate that Yukawa couplings
might acquire a complex phase of a geometric nature, quite apart
from the moduli dependence of these couplings, that can be
interpreted as an analog to nontrivial Wilson lines for gauge
fields (thus relating a geometrical breaking to a spontaneous
breaking)~\cite{Bailin:1998yt}. Furthermore, early in the study of
Calabi-Yau spaces it was noted that some Calabi-Yau manifolds
might admit an antiholomorphic isometry that could imply (a
presumably explicit) violation of CP~\cite{Strominger:1985it}.

Thus it is possible that, prior to any question of moduli
stabilization and supersymmetry breaking, a theory of CP violation
and a delineation of the possible physical and allowable phase
structure of the effective theory can be formulated already at the
string level. Such a formulation would be extremely powerful,
either in disfavoring certain classes of models or suggesting
others as phenomenologically interesting. In constructions where
it can be conclusively demonstrated that CP violation can only
occur through spontaneous symmetry breaking in the low energy
field theory, then it might still be possible to segregate
contributions into those related to vevs for the lowest components
of chiral superfields (and thus likely to be related to effective
Yukawa couplings and a theory of flavor) and those related to vevs
for the auxiliary fields of the same multiplets (and thus likely
related to a theory of supersymmetry breaking and transmission).
Such a classification would be a giant step forward in the
understanding of CP violation in the observable world. Hence we
ask:\\

\textbf{Question 19:} What are the stringy ways in which complex
phases can enter the observable world and can these be related to
a theory of flavor or supersymmetry breaking?

\section{Dynamical (super)symmetry breaking}
\label{sec:susy}
%

\subsection{Identifying hidden (and sequestered) sectors}

A principle question of formal string theory is the issue of how
many supersymmetries exist in the four-dimensional effective field
theory after compactification. This is a topic amenable to string
theory consideration in that it involves the determination of the
holonomy group of the compact manifold -- a geometrical question.
But once the theory is compactified, the question of breaking any
{\em remaining~} supersymmetries is then typically relegated to
the domain of phenomenology in as much as a field theory
explanation (such as gaugino condensation) is often employed.
Thus, it is often said, string theory has little to add to the
fundamental question of supersymmetry breaking once
compactification has occurred.

This argument is overly pessimistic: any particular string
construction can be characterized by certain allowed internal
geometries and the specific moduli fields that parameterize
them.\footnote{By ``string construction'' we mean not solely one
of the five or six branches of the M-theory amoeba but something
more specific, such as one of those branches and a class of
compactifications to 4D. For example, weakly coupled heterotic
$E_8 \times E_8$ on orbifolds, strongly coupled heterotic $E_8
\times E_8$ with $D_5$-branes on a Calabi-Yau, Type-II string
theory with $D_6$-branes on orientifolds, etc.} In effective
theories derived from string constructions it is often precisely
the chiral superfields that represent these moduli that
communicate supersymmetry breaking to the fields of the
supersymmetric Standard Model. While the stabilization of these
moduli -- and hence the determination of the exact values of their
auxiliary fields -- continues to be a difficult problem most
profitably addressed by effective field theories, some key
features continue to be geometrical in nature. And geometrical
properties allow for more robust and generic statements that can
help separate classes of models.

The foremost of these features is the existence of ``hidden
sectors.'' Let us be precise in our definition of this often-used
term. A hidden sector, properly speaking, should be one that can
communicate to other sectors only via the supergravity multiplet
or non-gauge-charged (i.e. singlet) fields with no superpotential
of their own, which we will call moduli. If the supergravity
multiplet is the only source of communication we will call this
special class a sequestered sector model.  If gauge fields can
participate in the communication of supersymmetry breaking, then
this we will call this a ``partially hidden'' sector. It is still
hidden in the limited sense that a loop diagram is required to
communicate the supersymmetry breaking to the Standard Model
supermultiplets. But in some sense the introduction of gauge
messengers blurs the geometrical distinction between the sectors,
which is our primary interest. Note that direct tree-level
communication of supersymmetry breaking (``visible sector''
models) is not forbidden, but is usually not considered since in
that case one must then contend with the fact that with only the
Standard Model gauge group at least one squark or slepton must be
lighter than the corresponding
fermions~\cite{Csaki:1996ks,Kumar:2004yi}.

These distinctions are crucial to understanding the nature of SUSY
breaking transmission to the sector where the MSSM resides.
Low-energy models of SUSY-breaking (more correctly, models of the
{\em transmission~} of SUSY breaking to the MSSM) can be crudely
divided into those based on mediation by the superconformal
anomaly, those mediated by moduli and those mediated by gauge
fields. Sequestered sectors can only use anomaly mediation, hidden
sectors can use both anomaly mediation and moduli mediation
(somewhat misleadingly called ``gravity'' mediation in the
literature), and a partially hidden sector can use both of these
and gauge mediation, and perhaps other mechanisms as well. Given
the ubiquity of moduli in string theory a truly sequestered sector
may be hard to find. Some studies suggest that even when sectors
of the theory can be physically separated in the compact space by
a bulk which contains only the gravity multiplet, branes tend to
warp the internal geometry and induce (moduli-dependent)
supersymmetry breaking~\cite{Anisimov:2001zz,Anisimov:2002az}. If
this finding can be promoted to a theorem then it would be most
interesting for phenomenology in as much as the idea of anomaly
mediation, with its insensitivity to ultraviolet physics and its
automatic solution to the supersymmetric flavor problem, would be
strongly disfavored in string models.

But explicit string constructions of both the heterotic and
Type~II variety suggest that even hidden sectors may be rare, with
most being only partially hidden. For example, the $E_8 \times
E_8$ heterotic string has two sectors that are (in principle)
hidden from one another, while the $SO(32)$ heterotic string does
not -- but that is prior to compactification. Once the gauge group
is broken by Wilson lines the Abelian factors that arise upon
compactification tend to span both $E_8$ factors -- meaning these
sectors are now only partially hidden. The anomalous U(1) is a
prime example, and models have been constructed where the massive
gauge bosons of this anomalous U(1) play a role in transmitting
SUSY breaking to the observable
sector~\cite{Dvali:1996rj,Mohapatra:1996in}. Furthermore, states
which are bi-fundamental under a subgroup of the Standard Model
and under a hidden sector group have been shown to be present in
bosonic heterotic string
constrictions~\cite{Font:1989aj,Giedt:2001zw}, free-fermionic
constructions~\cite{Cleaver:1998gc}, Type~IIA
constructions~\cite{Cvetic:2002qa} and others.

This suggests that all three SUSY transmission mechanisms may be
present simultaneously, with the only question being which mode
dominates. This latter point is a statement about the relative
scales of symmetry breakdown in a hidden sector and the mass scale
of the suppression factor in the operator that connects this
breakdown to the Standard Model. The bottom line is that the issue
of how ``hidden'' the hidden sector truly is revolves around the
geometry of the manifold of compactification and its corollary:
the states of the low-energy massless spectrum. Even the scale of
the operators connecting various sectors in the low energy field
theory are set by the string construction being considered. Thus
we feel that it is not inappropriate to ask the following
question:\\

\noindent \textbf{Question 20:} Can compactifications be
considered and classified, at a topological level, so as to
identify those that give rise to sectors which are truly hidden
and/or sequestered from one another?

\subsection{Gaugino Masses}
Whether gaugino soft mass terms are degenerate or not is of great
phenomenological importance. For example, if the gaugino masses
are degenerate then the complex phases can be rotated away.
Moreover, if they are degenerate then light gluinos must imply
very light charginos and neutralinos (unless there is a
significant amount of extra matter at intermediate scales with
special properties), and the latter masses are very constrained by
LEP data. Electroweak symmetry breaking can only occur with tuned
soft masses in the soft supersymmetry breaking Lagrangian if
gaugino masses are degenerate. Finally, collider data is likely to
probe and measure the parameters of the gaugino sector of a
supersymmetric theory long before usable information on the scalar
masses or trilinear scalar couplings is available.

In heterotic string theories the tree level gaugino masses are
normally degenerate, in that they arise from the auxiliary field
of one unified chiral superfield: the dilaton. This property is
robust in the sense that it remains true over a wide array of
string theories and compactifications, though it need not hold in
some open string theories. However, even in the large class of
string theories with universal (tree level) gaugino masses these
masses can be suppressed -- e.g. if supersymmetry breaking is due
mainly to nonzero vevs for the auxiliary fields of various types
of moduli associated with the geometry of the compact space,
rather than that of the dilaton. Usually the one loop masses are
non-degenerate and relatively large when the tree level masses are
suppressed, so the resulting gaugino masses are not
degenerate~\cite{Binetruy:2000md}.

Although one is discussing soft masses that arise from
supersymmetry breaking, we think that one can answer such
questions without understanding the precise nature of how
supersymmetry is broken by considering classes of supersymmetry
breaking. For example supersymmetry breaking can be represented by
non-vanishing auxiliary fields for different classes of string
moduli. When combined with the moduli dependence of the gauge
kinetic function (known in many string constructions both at the
genus zero and genus one level) this yields a general structure
for gaugino masses in the low-energy theory. From this point the
issue of gaugino masses (at least at leading order) is related to
the question of gauge coupling unification (see
Section~\ref{sec:unification} above).\\

\noindent \textbf{Question 21:} In what classes of string theories
are tree level gaugino masses likely to be suppressed?

\subsection{The $\mu$-term ``opportunity''}
\label{sec:muterm}

One of the most important parameters in the low-energy
supersymmetric Lagrangian is the $\mu$ parameter: a supersymmetric
mass term which couples the two Higgs doublets of the MSSM. The
value and sign of this parameter strongly affects the neutralino
and chargino masses, sets the size for the mixing between scalar
fermion gauge eigenstates in the mass eigenstate basis and
constrains the possible minima of the electroweak Higgs potential.
If the value of this mass term was much larger than a TeV the
entire theory would be incompatible with observations. The fact
that string theory tends to predict that there are no fundamental
bilinears in the superpotential ({\em i.e.} there are no
supersymmetric mass terms for the massless spectrum) implies that
it must be generated dynamically somewhere below the string energy
scale. This is a major step forward in understanding why the $\mu$
parameter is so small.

There are two broad classes of solutions to completing this
understanding. One possibility is that the $\mu$ parameter is the
result of some dynamical symmetry breaking in which some field
$S$, which is a singlet under the gauge factors of the Standard
Model, couples through a superpotential term $W = \lambda S H_u
H_d$ to the Higgs doublets of the Standard Model. Such at term is
common in heterotic string constructions, for example, since the
$S$ field can be identified with the singlet under the
decomposition $E_6 \to SO(10)$ and the requisite coupling is
allowed in the $\mathbf{27}^3$ coupling of $E_6$. In many
circumstances the necessary vev occurs naturally at approximately
1~TeV through a radiative process not unlike that which drives
electroweak symmetry breaking in the
MSSM~\cite{Ellis:1985yc,Cvetic:1997ky}.

However, {\em a priori~} we have no reason to expect this
mechanism to explain why the $\mu$ parameter should be comparable
to the scale of supersymmetry breaking in the observable sector.
That it turns out this way often in the case described above
appears as a fortuitous accident. In the second class of
solutions, which has come to be called the Giudice-Masiero (GM)
mechanism~\cite{Giudice:1988yz}, this relation between scales is
automatic. If a term of the form
\begin{equation}
K = \frac{1}{2}\left[\alpha_{ij}(Z,\overline{Z}) \varphi^i
\varphi^j + \hc \right]
\label{GM} \end{equation}
appears in the K\"ahler potential then an effective supersymmetric
mass term will be generated when supersymmetry breaking occurs
proportional to the gravitino mass and the function
$\alpha_{ij}(Z,\overline{Z})$ plus its derivative.

The two Higgs doublets of the MSSM are unique in being the only
vector-like superfields of the MSSM field content. Nevertheless,
bilinears in the Higgs doublets are not the only guage-invariant
bilinears that can be constructed with the MSSM fields. If a
mechanism such as~(\ref{GM}) is the source of the $\mu$-term in
the low energy theory then there is likely to be some stringy
property of the fields identified as Higgs doublets that allows
this term (and only this term) to exist in the K\"ahler
potential.\footnote{The same is not strictly true of the first
$\mu$ term mechanism: we might also wish there to be a mass term
in the superpotential for right-handed neutrinos, and perhaps
other states, in the low energy theory.} To date there exists
precisely one explicit calculation of such terms in the K\"ahler
potential~\cite{Antoniadis:1994hg}, performed in the context of
heterotic string theory on a particular orbifold.

It is instructive to consider the peculiarities of this particular
case. The existence of a coupling such as~(\ref{GM}) required a
(2,2) compactification, so that the theory would contain states
that transform as both $\mathbf{27}$ and $\mathbf{\overline{27}}$
representations of $E_6$. This is already an interesting fact, as
such popular choices as the $Z_3$ orbifold do not fall into this
category. This suggests that the states we might wish to identity
as Higgs states of the MSSM are {\em not} those that arise from
the $\mathbf{10}$ representation of SO(10), as is usually assumed.
Furthermore, the structure of the function
$\alpha_{ij}(Z,\overline{Z})$ involved in the coupling~(\ref{GM})
suggests that in this case the doublets involved should be thought
of as Wilson line moduli, and that they should transform
non-trivially under certain $SL(2,Z)$ symmetries. As such, these
states will also appear in the stringy threshold corrections to
gauge kinetic functions at one loop in the effective field theory.
The important point is that a relation exists between the
existence of a desired coupling in the low-energy effective
supergravity Lagrangian and many other key properties of the
theory: the types of moduli present in the theory and their
couplings, the symmetries present in the low-energy theory, and
the relation between the string properties of various states and
their low-energy properties. This suggests that the
``$\mu$-problem'' really should be thought of as the
``$\mu$-opportunity,'' as this issue probes many features of the
underlying string theory:\\

\noindent \textbf{Question 22:} What are the requirements on the
fields that we wish to consider Higgs states in order to implement
the Giudice-Masiero mechanism to generate a $\mu$-term? \\

\noindent \textbf{Question 23:} Are there other string theory (as
opposed to field theory) mechanisms which guarantee the relation
$\mu \sim m_{3/2}$ while simultaneously predicting $\mu \to 0$ in
the supersymmetric limit?

\subsection{Majorana neutrino masses}
\label{sec:neutrino}

The solid experimental evidence of neutrino oscillations suggests
that neutrinos have finite masses. To explain the very small size
of these masses, relative to those of the other Standard Model
fermions, it is often assumed that right-handed neutrinos exist
and have a very large Majorana mass term, as well as a Yukawa
coupling to the lepton doublet and up-type Higgs
doublet~\cite{Gonzalez-Garcia:2002dz}. Such supersymmetric masses
are theoretically equivalent to the $\mu$-term of the Higgs
sector, only the challenge is greater. Now we must not only
understand why these masses are not at the Planck scale, but also
why they are not electroweak scale in size. Again, string theory
provides a partial answer to the first question. But as in the
case of the Higgs sector $\mu$ parameter these Majorana masses are
presumably generated by some dynamical mechanism.

Proper phenomenology requires $M_{\nu} \sim 10^{12} - 10^{16}
\GeV$ so the K\"ahler potential mechanism is unlikely to prove
useful. This leaves only the dynamical mechanism of generating
Majorana mass terms through the trilinear coupling of right-handed
neutrino bilinears to a Standard Model singlet which acquires a
vacuum value. Unlike the Higgs mass term of the previous question,
the standard see-saw mechanism demands that right-handed neutrinos
have a bilinear coupling in which each species of
chiral-superfield couples to itself -- a type of coupling that may
be forbidden by string-selection rules~\cite{Giedt:2005vx}.
Furthermore, the fields which we come to identify as right-handed
neutrinos may be linear combinations of fields which are Standard
Model singlets present at string energies, making the effective
Majorana mass matrix field-dependent and involved. In principle
such mixing between low-energy MSSM fields and string-energy
exotics could generate effective Majorana mass terms for Standard
Model non-singlets as well, which would generically be a
phenomenological disaster. Thus the string theory must select the
eventual right-handed neutrino superfield for special treatment:
\\

\noindent \textbf{Question 24:} If right-handed neutrinos are not
true singlets what are the string-theory properties of these
fields that make them the {\em only~} SM fields with a large
supersymmetric mass?

\section*{Conclusion}

The questions presented here are meant as a starting point for
further inquiry. It would be unfair to say that no answers yet
exist for some of these questions, but what we seek is deeper than
an ``existence proof'' that an answer is possible (though such a
first step is lacking in some of the questions we pose). Rather,
we are interested in broad statements that can be used to
distinguish string theories by their phenomenological properties
and which may lead to deeper understanding of how the formal
properties of string constructions manifest themselves in the
low-energy world of observations.

We are encouraged by the fact that so many of these questions lend
themselves to a geometrical interpretation at some stage in the
string construction. This suggests that a meaningful string
phenomenology can be built around the classification of string
theories by the properties of their moduli. To appreciate the
enormous potential for progress in this area, consider the
following. Moduli stabilization and spontaneous SUSY breaking and
transmission tend to come together in effective field theories
based on string models. This leads to a determination of the soft
supersymmetry breaking Lagrangian for the observable sector -- and
ultimately to an understanding of patterns of masses of
superpartners. But the moduli also appear in the Yukawa couplings
of these theories. Their symmetries are thus potentially flavor
symmetries and their vevs determine the magnitudes of these
couplings (though this is somewhat more complicated if some
``matter'' fields end up with vevs in the effective trilinear
couplings). This leads to a pattern of flavor textures and
eventually fermion masses. CP violation might enter the 4D world
via both SUSY breaking operators and Yukawa couplings which are
functions of these moduli, so it too emerges from this same
physics. Finally, one of the string moduli is an ideal candidate
for an inflaton and its (above-mentioned) couplings to SM fields
allows for reheating and possible baryogenesis (say through the
Affleck-Dine mechanism).

Is it a reasonable goal to imagine a theory that explains the
interlocking relationships between fermion masses, the nature of
dark matter, collider experiments and current cosmological
observations~(\textbf{Question 25})? Only string theory can ever
hope to provide such an all-encompassing theoretical framework.
More progress will occur in this exciting area if more people work
in the directions we suggest above. Our goal has been to focus
increased attention on some issues where better understanding will
lead to progress, and where the understanding of the theory may
have reached a level allowing answers to some questions of
phenomenological value.

\section*{Acknowledgements}
The authors would like to thank the Aspen Center for Physics for
their hospitality during the final stages of this work, as well as
Vijay~Balasubramanian, Joel~Giedt, Burt~Ovrut and many others for
their input and suggestions. B.N.~was supported by the
U.S.~Department of Energy under Grant No.~DOE-EY-76-02-3071.


\end{document}